\documentclass[twocolumn,showpacs,preprintnumbers,amsmath,amssymb]{revtex4}
\usepackage[dvips]{graphicx}% Include figure files
\usepackage{dcolumn}% Align table columns on decimal point
\usepackage{bm}% bold math

\begin{document}

%\preprint{APS/123-QED}

\title{Photoassociation spectroscopy of $^{88}$Sr: reconstruction of the wave function near the last node}
\author{Masami Yasuda,$^*$ Tetsuo Kishimoto, Masao Takamoto, and Hidetoshi Katori}
\affiliation{%
Engineering Research Institute, 
The University of Tokyo, Bunkyo-ku, Tokyo 113-8656, Japan
}%

\date{\today}% It is always \today, today,
           %  but any date may be explicitly specified
\begin{abstract}

We have investigated photoassociation (PA) spectra of ultracold $^{88}$Sr atoms near the $5s^2\ ^1\!S_0+5s5p\ ^1\!P_1$ atomic asymptote.
The intensity modulation of the PA lines was used to reconstruct the ground-state scattering wave function,
 whose last nodal point was determined to be $r_0=3.78(18)$ nm.
The PA lines also determined a precise lifetime of the $^1\!P_1$ state to be 5.263(4) ns. 
\end{abstract}

% PACS, the Physics and Astronomy
                             % Classification Scheme.	%\keywords{Suggested keywords}%Use showkeys class option if keyword
                              %display desired
\maketitle

%\section{\label{sec:level1}Introduction}
% Introduction, scattering length, its importance, measurement methods, background
Precise knowledge of atom-atom interactions is of significant importance in the creation and manipulation of ultracold atoms or molecules 
as well as in their application to precision measurements.
Investigations on light-induced atom losses in magneto-optical traps (MOTs), where collision processes are altered by the presence of near 
resonant photons~\cite{Sesko}, have triggered an emerging field of cold collisions with laser cooled atoms~\cite{PArefWeiner}.
The stability and dynamics of the Bose-Einstein condensates (BECs) are governed by the ultracold collisional properties~\cite{Fetter,Tiesinga, Ruprecht, Kagan}.
Their tunability via magnetic or optical method is utilized to produce ultracold molecules as well as molecular BECs~\cite{Regal, Herbig, Strecker, Zwierlein}.
From a metrological point of view, atomic collisions introduce unwanted frequency shifts that critically limit the performance of the 
state-of-the-art atomic fountain clocks~\cite{Sortais} and optical clocks~\cite{PTB2002}.

A two-body collision problem is most simplified for ultracold atoms. 
The ground-state wave function for the angular momentum $l=0$ ($s$-wave) state can be 
written as $R(r)=u_g(r)/r$ with $u_g(r)$ the solution of an ordinary one-dimensional (1D) Schr\"odinger equation and $r$ the interatomic separation.
The interaction is characterized by a single parameter, the $s$-wave scattering length $a$, which is associated with the last node of the scattering wave function $u_g(r)$ \cite{RefApp1}.
One of the most versatile experimental techniques for probing the wave function, especially its last node, is the photoassociation (PA) spectroscopy~\cite{PArefThorsheim} 
that measures the intensity modulation of the PA lines ~\cite{Node2a1, Node2a2, Node2a3}.
Since the demonstration of the technique in laser-cooled  Na~\cite{NaPAS} and Rb~\cite{RbPAS} atoms,
there have been extensive theoretical as well as experimental studies in 
alkali-metal systems \cite{PArefWeiner}. 
However, analysis of these PA lines are not easy because the hyperfine structure in their ground state as well as the excited state 
gives rise to complicated molecular potentials often referred to as {\it hyperfine spaghetti} \cite{Spaghetti}.
In contrast, alkaline-earth-metal system offers the simplest energy structures, i.e., the $^1\!S_0$ ground state and the $^1\!P_1$ excited state, 
which allows straightforward interpretation of PA spectra.
Recently a PA line profile was obtained for Ca, which determined possible range of the scattering length~\cite{CaPAS}.
A predissociation process was revealed through the broadening of the PA line profiles in Yb atoms~\cite{YbPAS}.
PA spectra near the dissociation limit was investigated for Sr, which determined the $^1\!P_1$ excited state lifetime~\cite{SrPAS}.

% Sales points of this paper
In this Rapid Communication we report on the determination of the last node of the scattering wave function for the $5s^2\ ^1\!S_0$ state
of $^{88}$Sr atoms by reconstructing the wave function through PA spectroscopy.
The PA spectra were investigated for detunings up to $\Delta=-600$ GHz below the $5s^2\ ^1\!S_0+5s5p\ ^1\!P_1$ atomic asymptote.
Thanks to a simple electronic structure of alkaline-earth-metal atoms, the intensity modulation of the 
PA spectra straightforwardly inferred the squared scattering wave function for the interatomic separation of $3.2-11$~nm. 
The last node of the scattering wave function was found to be $r_0=3.78(18)$ nm, which suggested a possible range for the scattering length $a$.
The PA resonances were also used to derive the most precise lifetime of 5.263(4) ns yet reported for the $5s5p\ ^1\!P_1$ state,
which will provide a stringent test for a many-body calculation of the atomic structures \cite{C6Stanton, C6Porsev, C6Mitroy}.

\begin{figure}
\includegraphics[width=0.9\linewidth]{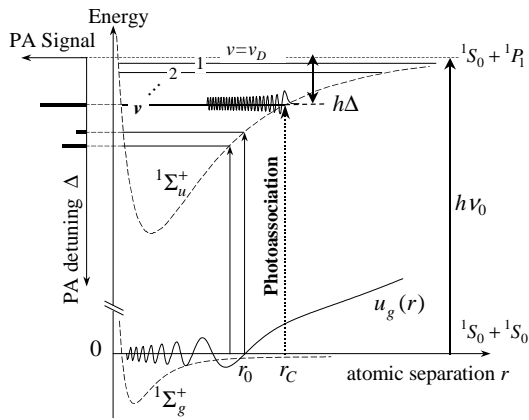}
\caption{Relevant Sr$_2$ molecular energy levels: The $^1\Sigma^+_g$ ground state and the attractive $^1\Sigma^+_u$ molecular states near the $5s^2\ ^1\!S_0+5s5p\ ^1\!P_1$ atomic asymptote  are shown by dashed lines.
At the Condon radius $r_C$, the PA laser excites an atom pair in the ground state to the $v$th vibrational levels of the $^1\Sigma^+_u$ state with a binding energy of $h\Delta$.
The PA signal intensity, as schematically depicted in the left panel, probes the squared amplitude of the scattering wave function $u_g(r)$ (solid line) to determine its nodal point of $r_0$.
}
\label{fig:MolecularPotential}
\end{figure}

% Explanation of PAS
We consider an atom pair in the $^1\!S_0$ ground state with relatively large interatomic distance $r$, where the resonant dipole-dipole (RDD) interaction 
dominates the atomic interaction. Figure~\ref{fig:MolecularPotential} shows schematic energy levels.
By shining a PA laser with frequency $\nu_L$, which is negatively detuned ($\Delta=\nu_L-\nu_0 < 0$) in respect to the $^1\!S_0-^1\!P_1$ atomic transition,
 the atom pair is excited to the $^1\!\Sigma^+_u$ molecular state bound by the RDD interaction. The molecular potential is given by
\begin{equation}
V_e(r)=D-\frac{C_3}{r^3},\ \ \ C_3=\frac{3 \hbar \lambdabar_0^3}{2 \tau_0},
\label{Eq:dipint}
\end{equation}
for $r < \lambdabar_0$ where the retardation effects \cite{Retardation} are negligible.
Here $D=h \nu_0$ is the energy of the $^1\!S_0 +{} ^1\!P_1$ atomic asymptote, and $\hbar=h/2 \pi$ is Planck's constant. 
The coupling constant $C_3$ for the $^1 \Sigma ^+ _u$ molecular state is expressed in terms of the reduced wavelength $\lambdabar_0 =\lambda_0/2\pi$ of the 
corresponding atomic transition and the radiative lifetime $\tau_0$ of the $^1\!P_1$ state. 
In the following discussion, we focus on atomic separations where the van der Waals interaction energy $V_g(r)$ is much less than the excited state width of $\sim \hbar \gamma_0$,
where $\gamma_0=1/\tau_0$ is the atomic linewidth.
The resonant excitation condition for the molecular potential is, therefore, given by
\begin{equation}
h \Delta=V_e(r)-V_g(r)-h \nu_0 \approx -C_3/r^3.
\label{Eq:rescond}
\end{equation}
In the spontaneous dissociation of the molecular states, the atom pair will acquire enough kinetic energy from the molecular potential through radiative decay
 or state changing processes \cite{Sesko} to be ejected from a shallow trapping potential. These atom-loss mechanisms are used to detect the PA process.

%\section{\label{sec:level3}Measurement}
The PA spectroscopy was done for a few microkelvins cold $^{88}$Sr atoms trapped in a 1D optical lattice~\cite{IdoLattice}, 
which was loaded from a narrow-line MOT operated on the $^1\!S_0-{}^3\!P_1$ transition at $\lambda=689$~nm with a linewidth of 7.5~kHz.
The 1D optical lattice consists of a standing wave of a far-off-resonant trap laser with a wavelength of $\lambda_L \approx 810$ nm.
 Its power of 250 mW was focused to the $e^{-2}$ beam radius of 
25 $\mu$m, which gave radial and axial trap frequencies of 350 Hz and 40 kHz, respectively.
With this trapping laser parameters, the population of the $^1\!P_1$ state, which is the most strongly coupled to the ground state by the trap laser, is approximately $10^{-9}$ and was neglected.
In addition, the induced light shift of $\delta U/h \approx$ 150 kHz and the Doppler shifts (motional sidebands) were much less than the molecular linewidth of $\gamma_0/\pi \approx 60$ MHz.
The typical number of atoms $N_0$ loaded into the optical lattice was $10^5$, corresponding to a peak density of $n_0\approx 10^{12}\ \mathrm{cm}^{-3}$. 
This high atomic density offered us a clear PA signal, as the two-body atom loss rate $\beta n_0$ could be made larger than the linear loss rate $\Gamma$ due to background gas collisions.
The ultracold temperature of atoms suppresses the higher order partial waves and a single rotational level ($J=1$) is allowed for the $^1\Sigma_u^+$ state.

The PA laser at $\lambda_0=461$~nm was generated by frequency doubling of a frequency-stabilized titanium sapphire laser (Coherent, {\small MBR110}), in a 7-mm-long waveguide periodically poled lithium niobate (PPLN) crystal. 
The mode-matched infrared power of 200~mW was converted to the blue output power of 70~mW.  
The PA beam with its $e^{-2}$ radius of 120 $\mu$m was superimposed on the 1D lattice beam.
Its intensity $I_{\mathrm{PA}}$ was typically 10 W/cm$^2$, which is
much larger than the saturation intensity $I_{\mathrm{sat}}=42$ mW/cm$^2$ of the associated atomic transition.
Such a rather intense PA laser was needed to excite deeply bound vibrational levels with a binding energy of $h \Delta$, since the Franck-Condon factor decreases as $|\Delta|^{-1/2}$ \cite{FCDep}.
Because of this $\Delta$-dependent PA rate, we set a duration $\tau_{\mathrm{PA}}$ of the PA laser 
in the range of $5-600$~ms so as to obtain typical atom losses of 50\% during a frequency scan.
The PA laser frequency was monitored by a wave meter (Burleigh, {\small WA-1500}), which was calibrated by the absorption spectrum of Te$_2$~\cite{TeSpectrumTable}.
One cycle of the measurement took 1 s, and the sweep rate of the PA laser frequency was 40 MHz/s,
which was continuously swept over 48 GHz in a single scan.    
%\section{Result and analysis}\

\begin{figure}
\includegraphics[width=0.9\linewidth]{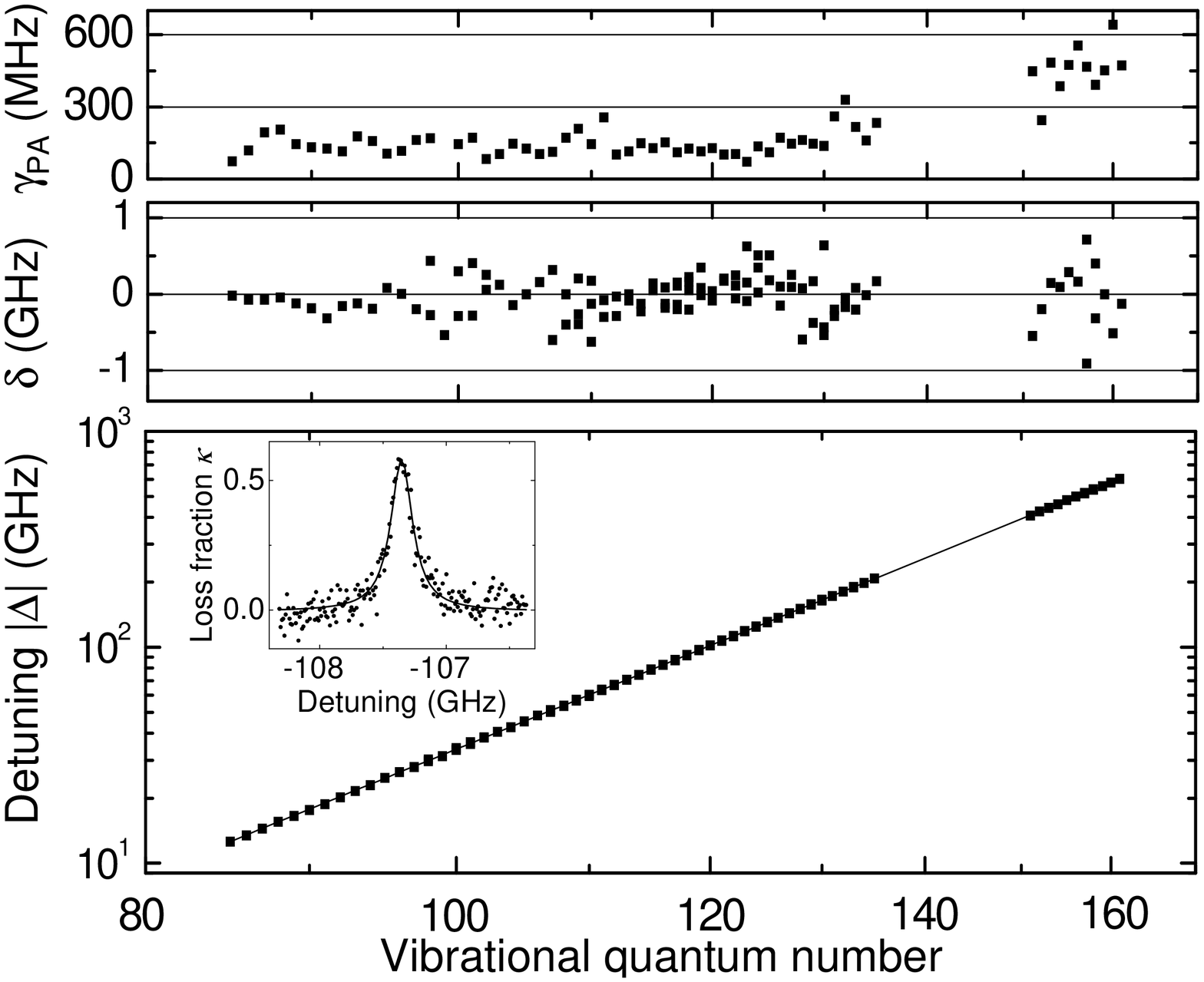}
\caption{PA resonances $|\Delta_{\mathrm{PA}}(v)|$ as a function of the vibrational quantum numbers
$v$ plotted on a log-log scale are shown in the lower panel.
 The solid line is the fit $|\Delta(v)|$ by the semiclassical formula given in Eq.~(\ref{Eq:DetuningVibnumber}).  
 The inset shows a typical PA line; atom loss fraction $\kappa$ as a function of detuning $\Delta$. 
The residuals of the fit, $\delta(v) = \Delta_{\mathrm{PA}}(v)-\Delta(v)$, are plotted in the middle panel.
The upper panel shows the PA linewidth $\gamma_{\mathrm{PA}}$. }
\label{fig:fitting}
\end{figure}

The inset of Fig.~\ref{fig:fitting} shows a typical PA atom-loss spectrum at $\Delta \approx -107$~GHz with $I_{\mathrm{PA}}=10$ W/cm$^2$ and $\tau_{\mathrm{PA}}=15$ ms.
A loss fraction of $\kappa(\Delta,\tau_{\mathrm{PA}})=[N_0-N(\tau_{\mathrm{PA}})]/N_0 \approx$ 0.5 was obtained, where $N_0$ and $N(\tau_{\mathrm{PA}})$ are the number of atoms before and after the PA process, 
respectively, measured by laser induced fluorescence on the $^1\!S_0-{}^1\!P_1$ cyclic transition. 
This PA linewidth $\gamma_{\rm PA}$ of a few hundred MHz was explained in terms of the power broadening, 
as it was decreased down to $(84\pm 12)$ MHz, which is roughly a molecular linewidth of $\gamma_0/\pi \approx 60$ MHz, by reducing the PA laser intensity to 40 mW/cm$^2$.
By increasing the detuning $|\Delta|$ of the PA laser, the PA signal $\kappa$ decreased faster than $|\Delta|^{-1/2}$ and was finally buried in the shot-to-shot atom number fluctuation of $\approx$10\% at $\Delta \approx -250$ GHz. 
We attributed this disappearance of the PA signal to the node of the ground-state wave function.
The vibrational lines reappeared at $\Delta \approx -400$ GHz and the PA signal $\kappa$ increased with further detuning. 
The upper panel of Fig.~\ref{fig:fitting} shows the change of the PA linewidth for $I_{\mathrm{PA}}=10$ W/cm$^2$. A line broadening started from $\Delta \approx -400$ GHz, which may be attributed to predissociation ~\cite{YbPAS}.
	
% Fitting to the semiclassical formula -> radiative lifetime.
We first analyze the observed PA resonances to confirm the validity of assuming the RDD interactions of Eq.~(\ref{Eq:dipint}) in our scanning range of $-600\,{\rm GHz}< \Delta < - 12\, {\rm GHz}$.
For this interaction potential close to the dissociation limit,
the vibrational energy $E(v)$ can be given by the semiclassical formula~\cite{LeRoy},
\begin{equation}
E(v)=D-X_0(v-v_D)^6,
\label{Eq:SemiclassicalEnergy}
\end{equation}
with
\begin{equation}
X_0=\frac{h^6}{\mu^3 C_3^2}\left[\frac{\Gamma(4/3)}{2\sqrt{2\pi}\Gamma(5/6)}\right]^6.
\label{Eq:VibrationalConstant}
\end{equation}
Here $v$ denotes the vibrational quantum number counted from the dissociation limit (the highest bound state is $v=1$), 
$0\le v_D <1$ is the effective quantum number of the dissociation limit, and $\mu$ the reduced mass.   
Figure~\ref{fig:fitting} summarizes the observed PA lines $|\Delta_{\mathrm{PA}}(v)|$ as a function of the vibrational quantum number $v$.
These resonances were fitted by 
\begin{equation}
\Delta(v) = -X_0 (v-v_D)^6/h,
\label{Eq:DetuningVibnumber}
\end{equation} 
with $X_0$ and $v_D$ being the fit parameters. 
The middle panel shows residuals $\delta(v) = \Delta_{\mathrm{PA}}(v)-\Delta(v)$ of the fit. The residuals $\delta(v)$ were comparable to the measurement uncertainty of the 
PA laser frequency of $\approx$ 300 MHz. 
Any systematic deviation of the fit was not found, which confirmed the validity of Eq.~(\ref{Eq:rescond}) that neglected the higher order dispersion terms such as $C_6$ and $C_8$.
The $X_0$ obtained by the fitting determined the coupling coefficient $C_3=(18.360\pm0.013)$ (atomic unit).
This inferred the atomic radiative lifetime of the $^1\!P_1$ state to be $\tau_0=(5.263\pm 0.004$) ns or the atomic linewidth of $\gamma_0/2 \pi=(30.24\pm 0.02)$ MHz, 
which is one order of magnitude improvement in uncertainty compared with a previous measurement \cite{SrPAS}.

The intensity modulation of the PA signal gives information on the amplitude of the ground-state scattering wave function $u_g(r)$, 
as the PA signal intensity is associated with the free-bound Franck-Condon (FC) factor $F(\Delta)$.
Following the reflection approximation \cite{RefApp1}, the FC factor is calculated by the squared ground-state scattering wave function evaluated at the Condon radius $r_C$,
\begin{equation}
F(\Delta)=\frac{\partial E(v)}{\partial v}\frac{1}{d_C}\left|u_g(r_C)\right|^2,
\label{eqn:FCF}
\end{equation}
where $d_C=\left|(d/dr)\left[V_e(r)-V_g(r)\right]\right|_{r=r_C}$ is the slope of the differential potential at $r_C$.
Neglecting the ground-state van der Waals potential, which is as small as $V_g/h \approx -200$ kHz even at the Condon radius corresponding to $\Delta = -600$ GHz, 
the Condon radius for the $v$th vibrational state is calculated as $r_C(v)=\sqrt[3]{C_3/[h\left|\Delta(v)\right|]}$ using Eq.~(\ref{Eq:rescond}).
Assuming the PA loss rate coefficient $\beta(\Delta)$ is proportional to the FC factor, Eq.~(\ref{eqn:FCF}) gives the squared ground-state wave function as
\begin{equation}
\left|u_g(r_C)\right|^2 \propto \frac{d_C}{\frac{\partial E(v)}{\partial v}} \beta(\Delta),
\label{eqn:Psi2}
\end{equation} 
which indicates that the scattering wave function can be reconstructed by the PA loss rate sampled at the Condon radius.

\begin{figure}
\includegraphics[width=0.9\linewidth]{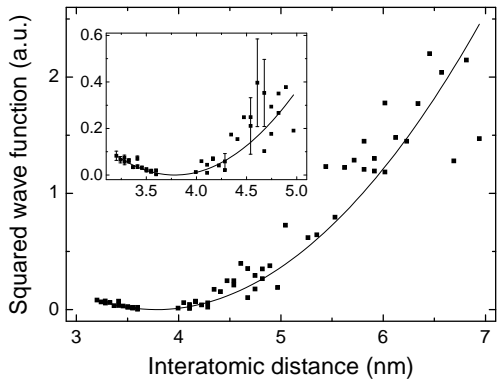}
\caption{The squared ground-state wave function $|u_g(r)|^2$ as a function of the interatomic distance $r$.
Filled squares indicate reconstructed wave function from the PA signals.
The inset is the enlarged view around the node, where the data points were used to be fit by 
$\left|u_g(r)\right|^2=A(r-r_0)^2$ (solid line) to determine the node to be $r_0=(3.78\pm 0.18)$ nm.}
\label{fig:FCF}
\end{figure}

The PA loss rate coefficient $\beta(\Delta)$ is determined from the PA loss fraction $\kappa(\Delta,\tau_{\mathrm{PA}})$ as
\begin{equation}
\beta(\Delta)=\frac{\Gamma}{n_0}\frac{\kappa(\Delta,\tau_{\mathrm{PA}})/\eta(\tau_{\mathrm{PA}})-1}{1-\kappa(\Delta,\tau_{\mathrm{PA}})},
\label{eqn:beta}
\end{equation}
by solving the rate equation for a time-dependent atom density $n(t)$ in the presence of collisions, i.e.,
$\dot{n}=-\Gamma n-\beta(\Delta) n^2$. 
Here $n_0$ is the initial atomic density, $\Gamma=(0.38\pm 0.05)$ s$^{-1}$ is the linear trap decay rate due to background gas collisions,
and $\eta(\tau_{\mathrm{PA}})=1-e^{-\Gamma \tau_{\mathrm{PA}}}$ is the linear loss fraction.
Figure \ref{fig:FCF} shows the reconstructed wave function $\left|u_g(r)\right|^2$ from the PA line intensities using Eq.~(\ref{eqn:Psi2}).
In this way, the disappearance of the PA lines is related to the nodal point of the scattering wave function.
Assuming the ground-state scattering wave function to be linear around the node $r=r_0$, i.e., $u_g(r)\propto (r-r_0)$,
 the squared wave function is given by 
\begin{equation}
 \left|u_g(r)\right|^2=A(r-r_0)^2,
 \label{eqn:fit}
\end{equation}
with $A$ a constant. 
The experimentally derived squared wave function as shown by dots in Fig.~\ref{fig:FCF} was fit by the above function, where we weighted the data points by their uncertainty.
The node of the ground-state wave function was determined to be $r_0=(3.78\pm0.18)$ nm.
Our previous measurements on the rethermalization time for ultracold $^{88}$Sr atoms \cite{Ido} inferred that the scattering length is not larger than 5 nm, 
or the last node to be $r_0<5$ nm, which suggests that the observed node is the last one.

% Observed node position to the scattering length (not so easy) 
The last node position $r_0$ can be straightforwardly associated with the $s$-wave scattering length $a$, especially when the node position is larger than a
quantum threshold radius of $r_Q$ \cite{R_Q} as in the case of Na \cite{Node2a3}.
For $^{88}$Sr, however, the observed node position was slightly smaller than this characteristic distance of $r_Q=4.0$~nm, therefore, a naive usage of the semiclassical formulas
 for calculating the scattering length from the node may be problematic. For example, the semiclassical formula Eq.~(11) in Ref.~\cite{Node2a3} inferred a positive scattering length 
of $a=(1.3\pm0.5)$ nm, while a quantum formula Eq.~(14) in the same reference gives a negative scattering length of $a=-2^{+2}_{-5}$ nm \cite{C6}.

%\section{Conclusion}
In conclusion, we investigated the PA spectra of ultracold $^{88}$Sr for detunings up to 600 GHz to the red of the dissociation limit. 
The observed PA lines determined the most precise natural lifetime of the $5s5p\ ^1\!P_1$ state of $^{88}$Sr yet reported.
The disappearance of the PA spectra gave us information on the nodal point of the ground-state wave function. 
The node, unfortunately, did not lead to a unique determination of the $s$-wave scattering length in the framework of the semiclassical treatment and urged careful quantum treatment.
However, small positive or negative scattering length suggested by this work indicates the difficulty in attaining BEC through evaporative cooling in $^{88}$Sr.

Recently, we became aware of Ref.~\cite{SrPASKillian} which measured the last node position of the ground-state $^{88}$Sr collision wave function.

%\section{Acknowledgement}
The authors thank S. Yamauchi for the technical support. This work received support from the Grant-in-Aid for Young Scientists (B) KAKENHI 15740254, the
 Japan Society for the Promotion of Science, and the Strategic Information and Communications R\&D Promotion Programme (SCOPE) of the Ministry of Internal Affairs and Communications of Japan.


\begin{thebibliography}{1}
\bibitem{FCDep}
		From Eq. (3) and (4), $\frac{\partial E(v)}{\partial v} \propto \left|\Delta \right|^{5/6}$. From the definition of $d_C$ and Eq. (2), it is shown that $d_C \propto \left|\Delta \right|^{4/3} $. 
Therefore, $F(\Delta) \propto \left|\Delta \right|^{5/6} \left|\Delta \right|^{-4/3}=\left|\Delta \right|^{-1/2}$ assuming the ground-state wave function to be constant.
\bibitem{C6}
		In calculating the scattering length, we take $C_6=(3210 \pm 40)$ (atomic unit), which is an average of three theoretical predictions in Refs. \cite{C6Stanton, C6Porsev, C6Mitroy}.
\end{thebibliography}
\end{document}